\documentclass[twocolumn,superscriptaddress,showpacs]{revtex4}
\usepackage{graphicx}

\begin{document}
\title{Harnessing quadratic optical response of two-dimensional materials through active microcavities}

\author{Alessandro Ciattoni}
\affiliation{Consiglio Nazionale delle Ricerche, CNR-SPIN, Via Vetoio 10, 67100 L'Aquila, Italy}
\author{Carlo Rizza}
\affiliation{Dipartimento di Scienza e Alta Tecnologia, Universit\`a dell'Insubria, Via Valleggio 11, 22100 Como, Italy}
\affiliation{Consiglio Nazionale delle Ricerche, CNR-SPIN, Via Vetoio 10, 67100 L'Aquila, Italy}

\begin{abstract}
We propose a method for efficiently harnessing the quadratic optical response of two-dimensional graphene-like materials by theoretically investigating second harmonic generation from a current biased sheet placed within a planar active microcavity. We show that, by tuning the cavity to resonate at the second harmonic frequency, a highly efficient frequency doubling process is achieved (several order of magnitude more efficient than the free-standing sheet). The efficiency of the process is not due to phase-matching, which is forbidden by the localization of the nonlinear quadratic response on the two-dimensional atomic layered material, but it stems from the interplay between the two-dimensional planar geometry of the nonlinear medium and the field oscillation within the active cavity near its threshold. The suggested method can easily be extended to different waves interactions and nonlinearities and therefore it can represent a basic tool for efficiently exploiting nonlinear optical properties of two-dimensional materials.
\end{abstract}


\maketitle

Two-dimensional materials and their prototype, graphene, have proved to be particularly suitable for optoelectronic applications \cite{Bonacc,BaoLoh} mainly since the Dirac cones characterizing their electronic band structures yield large carrier mobility and broadband light-coupling \cite{Staube}. Such resonant optical interaction also provides graphene with a large and broadband third order optical nonlinearity \cite{Hendry,ZhangV}
whose effect have been observed \cite{HongDa}. Since graphene is a centrosymmetric material, nonlinear effects due to second order optical nonlinearity are generally forbidden unless the space inversion symmetry is broken \cite{Margul} and biasing the sample with a direct current injection has been shown to be an efficient technique for observing second harmonic generation \cite{BykovM,AnNels}. Particulary interesting is current-biased bilayer graphene which, due to its four-band electronic structure and widely tunable bandgap in the mid-infrared, has been shown to exhibit a marked quantum-enhanced and tunable quadratic response \cite{WuMaoJ}.

Even though graphene-like materials have nonlinear properties much stronger than bulk materials, exploiting the ensuing nonlinear optical effects for conceiving light steering devices is hampered by their very small thickness.  A possible way for overcoming such limitations is to resort to field enhancement mechanisms \cite{CaoHal,Mondia,Siltan,Vincen,Ciatto} which have to be compatible with the planar geometry of two-dimensional materials. Recently, strong field enhancement effects have been considered in the presence of graphene \cite{GanMak,Thongr}. However, to the best of our knowledge, achieving feasible nonlinear light steering by combining graphene optical nonlinearity with a field enhancement mechanism has been shown in a single paper by Gu {\it et al.} \cite{GuPetr} where the authors show that placing a graphene sheet on the top of a silicon photonic crystal hosting a high-Q cavity (responsible for a large in-cavity field enhancement) produces optical bistability, self induced regenerative oscillations and coherent four-wave mixing at ultra-low optical intensities.

Observing strong effects due to quadratic optical response of two-dimensional materials is a even a more challenging task since the small sheet thickness forbids phase matching. In this Letter we show that the quadratic optical response of two-dimensional materials can efficiently be harnessed by placing the current-biased sheet within a planar active cavity which is tuned to resonate at the second harmonic frequency. We point out that the field enhancement produced by the gain amplification is compatible with the planar localization of the quadratic response and such cooperation produces a very efficient frequency doubling process. We find that the second harmonic generation efficiency is, in our scheme, several order of magnitude greater than that of the free-standing sheet. Since the proposed scheme can be extended even to third order and more general planar nonlinearities and it is effective with any two-dimensional material, we believe that our approach could further increase the present already intense research interest on nonlinear properties of atomic layered materials and could pave the way for a novel generation of compact nonlinear light-steering devices.

\begin{figure}
\center
\includegraphics*[width=0.45\textwidth]{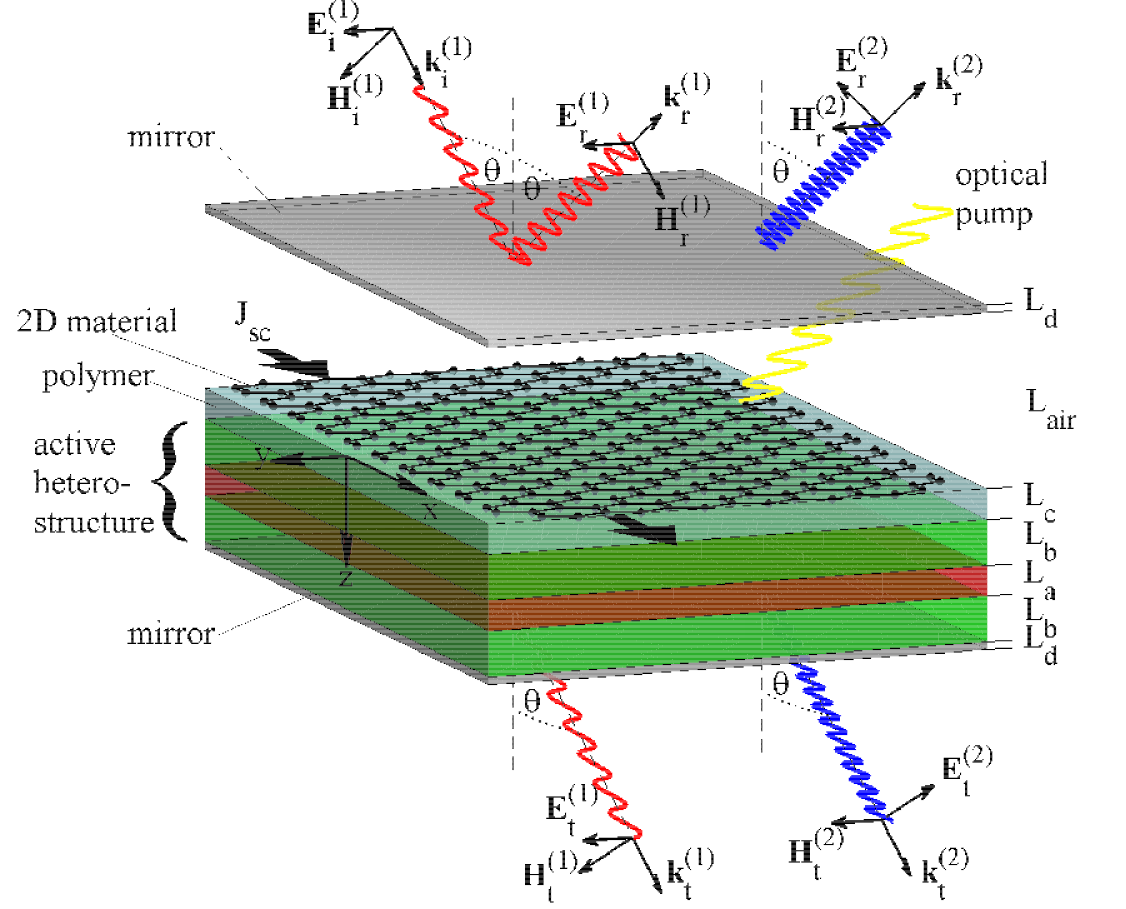}
\caption{(Color on-line). Geometry of the external cavity containing the current-biased two-dimensional material placed of the top of a polymer layer depositated onto an active heterostructure. The cavity is illuminated by an inclined transverse electric (TE) fundamental monochromatic wave (i) which produces reflected (r) and transmitted fundamental waves (t) together with reflected (r) and transmitted (t) transverse magnetic (TM) second harmonic waves. The cavity is also illuminated by an inverting pump beam.}
\end{figure}

In Fig.1 the setup is sketched together with the geometry of the fundamental and second harmonic waves (labelled with superscripts $(1)$ and $(2)$ respectively), the whole bichromatic electromagnetic field being ${\bf E}=Re \left[{\bf E}^{(1)} e^{-i \omega t} + {\bf E}^{(2)} e^{-i 2 \omega t} \right]$ and ${\bf H}=Re \left[{\bf H}^{(1)} e^{-i \omega t} + {\bf H}^{(2)} e^{-i 2 \omega t} \right]$. The considered external cavity between the two mirrors contains an air layer, an active semiconductor heterostructure and the two-dimensional atomic layered medium which is placed on the top of a polymer layer deposited on the heterostructure. The two-dimensional material is biased by a stationary (dc) surface current $J_{sc}$ flowing along the $x$-axis whereas the cavity is laterally optically pumped. The system is excited by a fundamental plane wave which is launched at an angle $\theta$ and consequently the cavity scatters the fundamental field and produces the second harmonic waves, both fields having transmitted (subscript $t$) and reflected (subscript $r$) components. In the following we will focus on the situation where the active cavity is tuned to resonate with the second harmonic field.

Due to the extremely small thickness of the two-dimensional material, we model its effect on the optical field through the matching conditions $\hat{\bf n} \times \left(\bf{E}^+ - \bf{E}^-\right) =0$ and $\hat{\bf n} \times \left(\bf{H}^+-\bf{H}^-\right)  = {\bf K}$, i.e. the continuity of the electric field tangential component and the discontinuity of the magnetic field tangential component produced by the surface current ${\bf K} = Re \left[{\bf K}^{(1)} e^{-i \omega t} + {\bf K}^{(2)} e^{-i 2 \omega t} \right]$. Since the space inversion symmetry of the two-dimensional material is broken by the biasing dc current, the surface current harmonics can be represented by \cite{WuMaoJ} $K^{(1)}_\alpha = \sigma^{(1)}_1 E^{(1)}_\alpha + [\sigma^{(1)}_2]_{\alpha \beta \gamma} E^{(1)*}_\beta E^{(2)}_\gamma$ and $K^{(2)}_\alpha = \sigma^{(2)}_1 E^{(2)}_\alpha + [\sigma^{(2)}_2]_{\alpha \beta \gamma} E^{(1)}_\beta E^{(1)}_\gamma$, where greek subscripts run over $x,y$. Here the first order conductivities $\sigma^{(n)}_1$ are negligibly affected by $J_{sc}$ and the second order conductivity tensor $\sigma^{(n)}_1$ and $\sigma^{(n)}_2$ strongly depend on $J_{sc}$. Since the surface current $J_{sc}$ is along the $x$-axis, it turns out that $[\sigma^{(n)}_2]_{xxy}= [\sigma^{(n)}_2]_{xyx} = [\sigma^{(n)}_2]_{yxx}=[\sigma^{(n)}_2]_{yyy} =0$ and therefore the field matching conditions imply that, if the fundamental field is transverse electric (TE with electric field along the $y$-axis) the produced second harmonic field is transverse magnetic (TM with magnetic field along the $y$-axis) as depicted in Fig.1. Accordingly, the field matching conditions at the plane $z=0$ read
\begin{eqnarray} \label{match}
E^{(1)+}_y - E^{(1)-}_y &=& 0, \nonumber \\
H^{(1)+}_x - H^{(1)-}_x &=& \sigma^{(1)}_1 E_y^{(1)+} + [\sigma^{(1)}_2]_{yyx} [E_y^{(1)+}]^* E_x^{(2)+}, \nonumber \\
E^{(2)+}_x - E^{(2)-}_x &=& 0, \nonumber \\
H^{(2)+}_y - H^{(2)-}_y &=& - \sigma^{(2)}_1 E_x^{(2)+} -[\sigma^{(2)}_2 ]_{xyy} [E_y^{(1)+}]^2.
\end{eqnarray}
The nonlinear terms appearing in the second and third of Eqs.(\ref{match}) are responsibile for the interaction of the two field harmonics in the scheme
we are considering.

The effect of the cavity is introduced by relating the fields on both sides of the two-dimensional material to the amplitudes $E^{(n)}_s$ ($n=1,2$ and $s=i,r,t$) of the plane waves outside the cavity. In the region $z<0$ the mirror and the air layers have a linear optical response and therefore we have $H^{(1)-}_x = q^{(1)}_{xi} E^{(1)}_i + q^{(1)}_{xr} E^{(1)}_r$, $E^{(1)-}_y = Q^{(1)}_{yi} E^{(1)}_i + Q^{(1)}_{yr} E^{(1)}_r$, $E^{(2)-}_x = Q^{(2)}_{xr} E^{(2)}_r$, $H^{(2)-}_y = q^{(2)}_{yr} E^{(2)}_r$ where all the factors $q$ and $Q$ are easily evaluated by means of the transfer matrix method and which depend on the physical and geometrical properties of the layers and on the incidence angle $\theta$. On the other hand, in the region $z>0$ all the layers have a linear optical response with the exception of the active layer (the layer with thickness $L_a$ in Fig.1) where the second harmonic field experiences nonlinear propagation due to gain saturation (the first harmonic is not resonant with the cavity and therefore linearly propagates everywhere). Specifically we assume the gain coefficient in the active layer to be described by the standard saturated model which is derived from the rate equations approach \cite{Suhara}, namely
\begin{equation} \label{gain}
g=\frac{B(I_p-I_{p0})}{1+I^{(2)}/I^{(2)}_s}
\end{equation}
where $I_{p}$ and $I_{p0}$ are the pump intensity and the transparency pump intensity, respectively, B is a coefficient depending on the medium differential gain, $I^{(2)}$ is the optical intensity of the second harmonic wave and $I^{(2)}_s$ is the saturation intensity. As a consequence, on the side $z=0^{+}$ of the two dimensional material, the boundary fields can be expressed as $H^{(1)+}_x = q^{(1)}_{xt} E^{(1)}_t$, $E^{(1)+}_y = Q^{(1)}_{yt} E^{(1)}_t$, $E^{(2)+}_x = F^{(2)}_{xt} (E^{(2)}_t)$ and $H^{(2)+}_y = f^{(2)}_{yt} (E^{(2)}_t)$, where the factors $q$ and $Q$ are again evaluated through the transfer matrix method whereas $F$ and $f$ are functions of $E^{(2)}_t$ which have to be evaluated by using the transfer matrix method for the linear layers and by solving Maxwell equations for the second harmonic field backward in the active layers using the dielectric permittivity $\epsilon = \epsilon_a -i \sqrt{\epsilon_a} \frac{\lambda^{(2)}}{2 \pi} g$ (where $\epsilon_a$ is the background dielectric permittivity a $\lambda^{(2)}$ the second harmonic wavelength). Inserting the boundary fields into the matching conditions of Eqs.(\ref{match}) and using the two equations for the continuity of the tangential electric field components to eliminate the reflected amplitudes one easily obtain the equations
\begin{eqnarray} \label{basic}
A^{(1)}_t E^{(1)}_t - [\sigma^{(1)}_2]_{yyx} [Q^{(1)}_{yt} E^{(1)}_t]^* F^{(2)}_{xt} - A^{(1)}_i E^{(1)}_i &=& 0, \nonumber \\
f^{(2)}_{yt} + a^{(2)}_t F^{(2)}_{xt} + [\sigma^{(2)}_2 ]_{xyy} [Q^{(1)}_{yt} E^{(1)}_t]^2 &=& 0 \nonumber \\
\end{eqnarray}
where $A^{(1)}_t = q^{(1)}_{xt} - q^{(1)}_{xr} Q^{(1)}_{yt}/Q^{(1)}_{yr} - \sigma^{(1)}_1 Q^{(1)}_{yt}$, $A^{(1)}_i = q^{(1)}_{xi} - q^{(1)}_{xr} Q^{(1)}_{yi}/Q^{(1)}_{yr}$ and $a^{(2)}_t = - q^{(2)}_{yr}/Q^{(2)}_{xr} + \sigma^{(2)}_1$ are geometrical factors characterizing the linear cavity. Equations (\ref{basic}) fully describe the considered frequency doubling process since, for a given $E^{(1)}_i$, they allow to evaluate $E^{(2)}_t$ (which is implicitly defined) and $E^{(1)}_t$. Note that $E^{(2)}_t$ does not vanish as a consequence of the third term in the second equation which is produced by the quadratic response of the two-dimensional material.

\begin{figure}[t]
\center
\includegraphics*[width=0.48\textwidth]{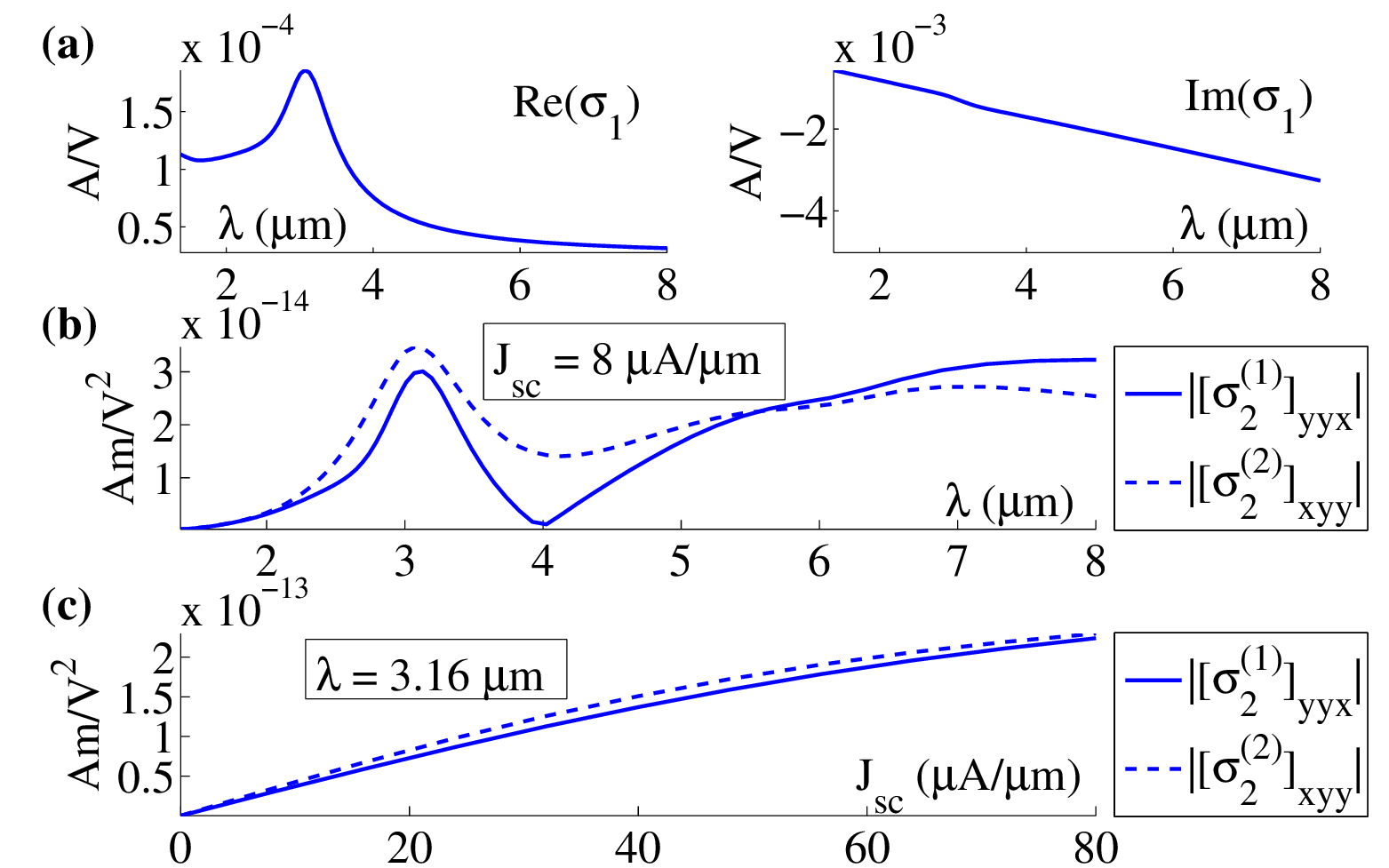}
\caption{(Color on-line). Room temperature linear and nonlinear optical properties of bilayer graphene in the mid-infrared. The chosen chemical potential is $\mu = 0.2 \: eV$. (a) Real and Imaginary part of the linear conductivity. (b) Relevant second order conductivity tensor coefficients evaluated for the biasing surface current $J_{sc}= 8 \: \mu A$. (c) Dependence of the second order conductivity tensor coefficients on the biasing surface current $J_{sc}$ at $\lambda = 3.16 \: \mu m$.}
\end{figure}

In order to physically grasp the interplay between the planar quadratic nonlinearity and the resonant cavity effects, let us consider the situation where the amplitude $E^{(1)}_i$ of the exciting fundamental wave is so small that the produced second harmonic field is such that, within the active layer,
$I^{(2)} \ll I^{(2)}_s$ so that the gain coefficient of Eq.(\ref{gain}) becomes $g \simeq B(I_p-I_{p0}$). Since gain no longer depends on $I^{(2)}$, the second harmonic wave linearly propagates through the active layer so that we have $F^{(2)}_{xt} (E^{(2)}_t) = Q^{(2)}_{xt} E^{(2)}_t$ and $f^{(2)}_{yt} (E^{(2)}_t) = q^{(2)}_{yt} E^{(2)}_t$  where the factors $Q$ and $q$ can again be easily evaluated through the transfer matrix method. At the same time, the second term in the first of Eqs.(\ref{basic}) can be neglected (undepleted pump condition) so that $E^{(1)}_t = (A^{(1)}_i/A^{(1)}_t) E^{(1)}_i$ which, inserted in the linearized version of the second of Eqs.(\ref{basic}), yields
\begin{equation} \label{linearized}
E^{(2)}_t = -\frac{[\sigma^{(2)}_2 ]_{xyy}}{A^{(2)}_t} \left[ Q^{(1)}_{yt}\frac{A^{(1)}_i}{A^{(1)}_t} E^{(1)}_i \right]^2
\end{equation}
where $A^{(2)}_t = q^{(2)}_{yt} + a^{(2)}_t Q^{(2)}_{xt}$ is a geometrical factor characterizing the linear cavity. Equation (\ref{linearized}) allows $E^{(2)}_t$ to be evaluated for a given $E^{(1)}_i$ and it is evident that, due to the absolute small value of $[\sigma^{(2)}_2 ]_{xyy}$, the amplitude of the produced second harmonic is generally very small. The scenario dramatically changes when $A^{(2)}_t$ is close to zero and this precisely occurs at the cavity lasing threshold for the second harmonic field. Therefore, by acting on the incidence angle $\theta$ and the pump intensity $I_{p}$, one can tune the cavity to resonate with the second harmonic field and, in view of Eq.(\ref{linearized}), this amounts to a powerful field enhancement mechanism yielding a strong second harmonic signal. Evidently when this happens the considered linearized scheme is no longer appropriate and gain saturation comes into play thus avoiding unphysical divergences. In this regime the situation is that a lasing cavity driven by an external field, the remarkable thing being that the driving field ($E^{(1)}_i$) has half the frequency of the field produced by the cavity ($E^{(2)}_t$) as a consequence of the frequency doubling process triggered by the two-dimensional material. The crucial point is here that the lasing cavity resonating with the second harmonic field is a very efficient mean for harnessing the two-dimensional material quadratic response which, without cavity, produces much lower (several order of magnitude) second harmonic signals (see below). It is worth stressing that the whole process is here made possible by the atomic thickness of the two-dimensional material since, for a bulk quadratic nonlinear medium placed within a lasing cavity, Eqs.(\ref{basic}) can not even be written.

We now discuss the above proposed strategy for harnessing the planar quadratic nonlinearity in a realistic setup. Specifically we have chosen bilayer graphene as the two-dimensional nonlinear material and, following the approach of Ref.\cite{WuMaoJ}, we have evaluated $\sigma^{(n)}_1$ and the tensors $\sigma^{(1)}_2$ and $\sigma^{(2)}_2$ (the former tensor has not been discussed in Ref.\cite{WuMaoJ}). The resultant optical properties are reported in Fig.2 where we have used the temperature $T=300 K$ and the chemical potential $\mu = 0.2 \: eV$. In Fig.2(a) the real and imaginary part of the linear conductivity are reported as a function of the wavelength and we have checked that $\sigma_1$ effectively does not depend on the biasing surface current $J_{sc}$. In Fig.2(b) the moduli of the relevant quadratic conductivity coefficients for $J_{sc}= 8 \: \mu A / \mu m$ are plotted as a function of the wavelength. In Fig.2(c) the moduli of the same coefficients are plotted for $\lambda = 3.16 \: \mu m$ as a function of the biasing surface current $J_{sc}$. To fully exploit such marked nonlinear features, we have set $\lambda^{(1)} = 3.16 \: \mu m$ and $\lambda^{(2)} = 1.58 \: \mu m$ for the vacuum fundamental and second harmonic wavelengths. In order to achieve resonance at $\lambda^{(2)}$ we have designed a cavity containing a $InP/ Ga_{0.47} In_{0.53} As /InP$ heterostructure with undoped layers (layers b/a/b of Fig.1) \cite{Miller} with a Polyethyl methacrylate (PMMA) substrate (layer c of Fig.1) for the bilayer graphene and silver mirrors (layers d of Fig.1).
\begin{figure}[t!]
\center
\includegraphics*[width=0.48\textwidth]{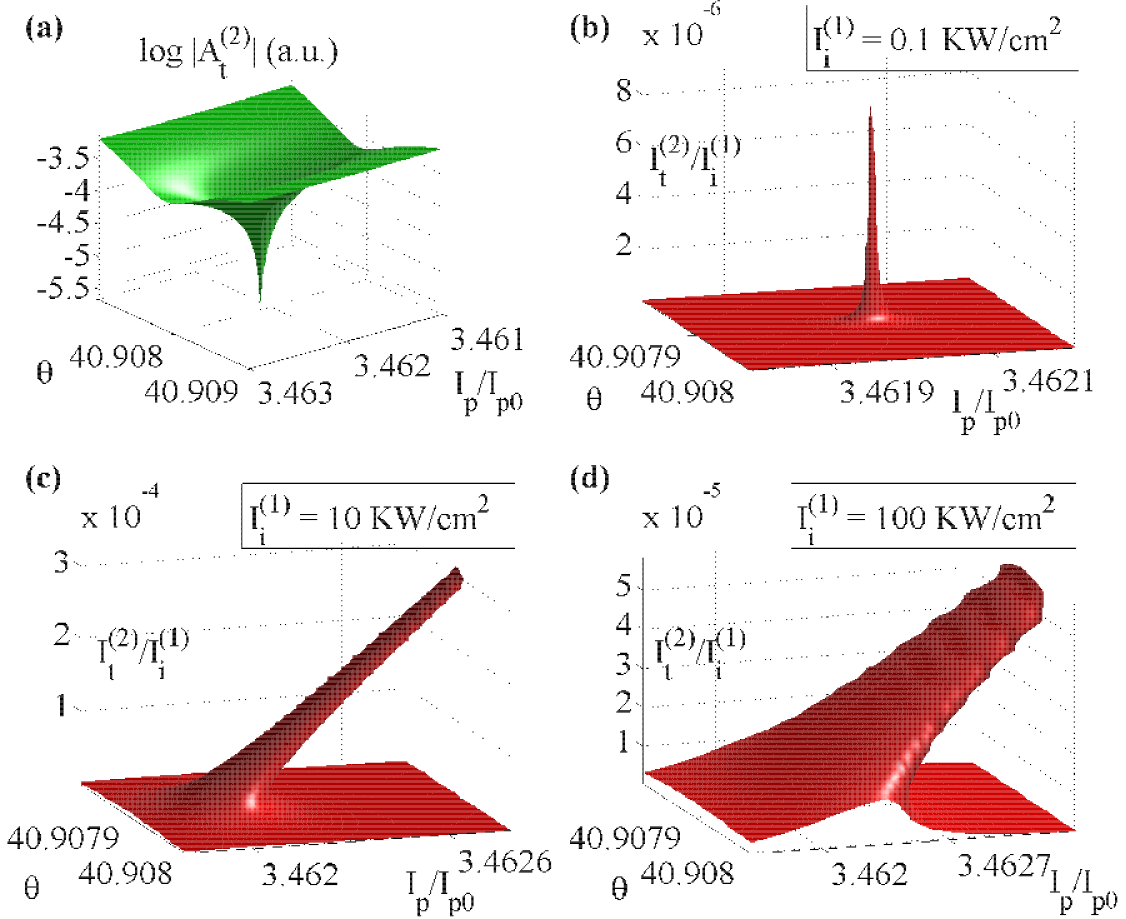}
\caption{(Color on-line) (a) Logarithmic plot of the quantity $|A^{(2)}_t|$ whose zeros identify the cavity lasing thresholds at $\lambda^{(2)}$. (b)(c)(d) Frequency doubling efficiencies $I_t^{(2)}/I_i^{(1)}$ as functions of the angle $\theta$ (degrees) and the normalized pump intensity $I_p/I_{p0}$ for different values of the intensity $I_i^{(1)}$ of the incident fundamental field.}
\end{figure}
We have set $L_a = 1.5 \: \mu m$, $L_b = 0.8 \: \mu m$, $L_c = 0.5 \: \mu m$, $L_d = 0.04 \: \mu m$ and we have used the well known physical properties of the considered semiconductor heterostructure \cite{AsadaY,Ahrenk} to obtain $B= 7.49 \: m/W$, $I_{p0}= 3.86 \: KW/cm^2$ and $I_s^{(2)} = 67.52 \: KW/cm^2$ as parameters appearing in Eq.(\ref{gain}). Besides we ha used the standard properties of PMMA and silver and the linear and quadratic response of bilayer graphene reported in Fig.2. In order to find the lasing threshold at $\lambda^{(2)}$ we have evaluated the quantity $A^{(2)}_t$ (see Eq.(\ref{linearized})) characterizing the cavity with unsaturated gain and we have plotted the logarithm of its modulus in Fig.3(a) from which it is evident that the threshold occurs for $\theta = 40.9079 \: deg$ and $I_p = 3.462 \: I_{p0}$. Subsequently we have solved the full boundary value problem of Eqs.(\ref{basic}), thus obtaining the second harmonic intensity $I_t^{(2)} = (1/2) \sqrt{\epsilon_0/\mu_0} |E_t^{(2)}|^2$ for different values of the fundamental wave intensity $I_i^{(1)} = (1/2) \sqrt{\epsilon_0/\mu_0} |E_i^{(1)}|^2$ and by varying, for each $I_i^{(1)}$, both $\theta$ and $I_p$ close to the threshold values.
\begin{figure}[tt]
\center
\includegraphics*[width=0.48\textwidth]{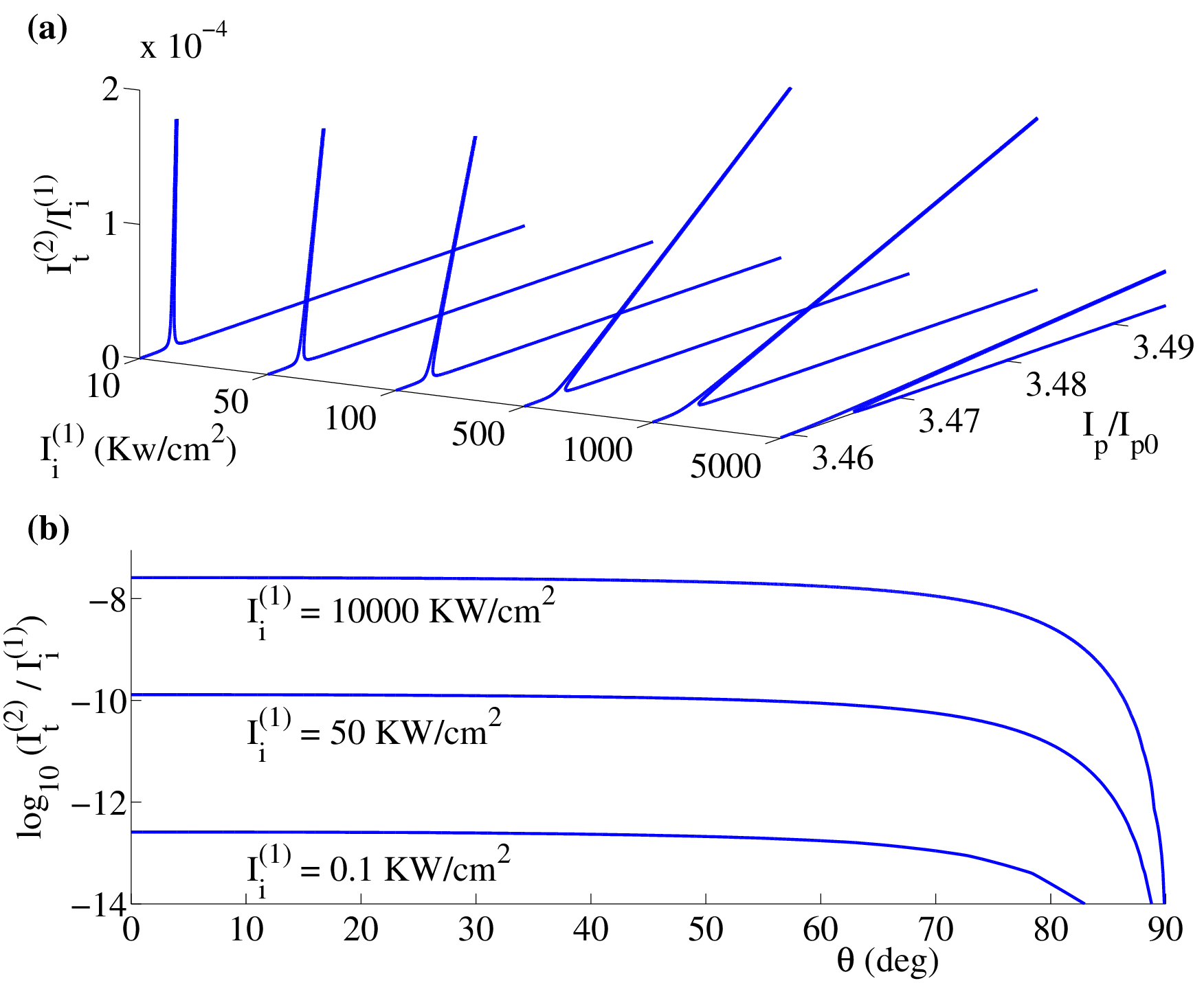}
\caption{(Color on-line) (a) Frequency doubling efficiencies $I_t^{(2)}/I_i^{(1)}$, for $\theta$ at is threshold value, as function of $I_p$ for different $I_i^{(1)}$. (b) Frequency doubling efficiency, as function of $\theta$, of a free-standing bilayer graphene sheet for various $I_i^{(1)}$.}
\end{figure}
In Figs.3(b), 3(c) and 3(d) we plot the frequency doubling efficiency $I_t^{(2)}/I_i^{(1)}$ for the intensities $I_i^{(1)} = 0.1, 10, 100 \: KW/cm^2$, respectively. Note that, as predicted, the second harmonic signal practically vanishes unless the incident angle $\theta$ is very close to its threshold value. As far as the optical pump intensity, note that, at small $I_i^{(1)}$ (Fig.3(b)), $I_p$ has to be very close to its threshold value to allow second harmonic generation whereas, at higher $I_i^{(1)}$ (Figs.3(c) and (d)), $I_p$ can assume even greater values. This occurs since, for small $I_i^{(1)}$, the produced second harmonic field is so weak within the cavity to forbid gain saturation and the scheme leading to Eq.(\ref{linearized}) actually works with the second harmonic field enhancement only occurring at the zero of $A^{(2)}_t$, i.e. at threshold (note that the peak of Fig.3(b) is located around the deep of Fig.3(a)). For higher values of the fundamental field intensity $I_i^{(1)}$, the second harmonic signal within the cavity is correspondingly higher and it triggers gain saturation allowing, as happens in the standard laser mechanism, signal amplification even for pump intensities $I_{p}$ greater than its threshold value. In Fig.4(a) we plot the frequency doubling efficiencies $I_t^{(2)}/I_i^{(1)}$, for $\theta$ at its threshold value, as function of $I_p$ for different values of $I_i^{(1)}$. Note that if $I_p$ is smaller than its threshold value no appreciable second harmonic generation occurs whereas, after the threshold, the intensity of the produced second harmonic field linearly increases with $I_p$, the slope being dependent on $I_i^{(1)}$.

In order to fully appreciate the reach of the proposed method for harnessing the quadratic nonlinearity of two-dimensional materials, we have evaluated the frequency doubling efficiency of a free-standing bilayer graphene sheet (biased by the current $J_{sc} = 8 \: \mu A / \mu m$ as above) for various $I_i^{(1)}$ and we have plotted the results in Fig.4(b). Note that, even at the higher chosen intensity $I_i^{(1)} = 10000 \: KW/cm^2$ the efficiency  $I_t^{(2)}/I_i^{(1)}$ is of the order of $10^{-8}$ which very much smaller than the efficiencies predicted in Figs.3(a), 3(b), 3(c) and 4(d). As an example, for $I_i^{(1)} = 0.1 \: KW/cm^2$ the free-standing sheet has a second harmonic generation efficiency of the order of $10^{-12}$ (see Fig.4(b)) which is insignificant if compared to $I_t^{(2)}/I_i^{(1)} \simeq 10^{-6}$ (see Fig.3(b)) occurring in the above proposed setup. We conclude that the proposed method effectively provides a strong enhancement mechanism which substitutes the unfeasible phase-matching coupling in the presence of two-dimensional materials.

In conclusion we have proposed a strategy to fully exploit the quadratic optical response potentials of two-dimensional graphene-like materials. The method is based on the combined effect of the planar localization of the nonlinear response and the amplifying properties of the hosting cavity. We have focused on second harmonic generation which turns out to be several order of magnitude more efficient than that due to a free-standing two-dimensional material sheet. A relevant feature of the proposed method is that it can be applied in a number of different situations and interaction geometries. For example, if the cavity is tuned to resonate with first harmonic wave, a very strong and different doubling frequency process is expected (since in this case $A_t^{(1)}$ in Eq.(\ref{linearized}) is very close to zero ). In addition an intriguing nonlinear interaction is expected if both the harmonics are launched due to the cavity-enhancement of their mutual energy exchange. Generalizing even more, the proposed method is expected to work also in the presence of the third order nonlinear response or different nonlinearities and it can be exploited in the whole frequency range where an amplifying cavity is available. In view of such generality and efficiency we believe that our approach can in principle be the platform to devising novel schemes and concepts where the marked nonlinear optical response of two-domensional material is efficiently exploited to achieve complex light manipulation and steering.


\end{document}